\documentclass{article}

\usepackage[overload]{textcase}
\usepackage{PRIMEarxiv}
\usepackage{tabularx}
\usepackage[table]{xcolor}
\usepackage{makecell}
\usepackage[utf8]{inputenc} 
\usepackage[T1]{fontenc}    
\usepackage{hyperref}       
\usepackage{url}            
\usepackage{booktabs}       
\usepackage{amsfonts}       
\usepackage{nicefrac}       
\usepackage{microtype}      
\usepackage{lipsum}
\usepackage{fancyhdr}       

\usepackage{graphicx}       
\usepackage{subcaption}
\graphicspath{{media/}}     
\usepackage{caption}
\usepackage{amsmath}
\title{\MakeLowercase{From First-Order to Higher-Order Interactions: Enhanced Representation of Homotopic Functional Connectivity through Control of Intervening Variables}
}

\author{
  Behdad Khodabandehloo\textsuperscript{a,*}, Payam Jannatdoust\textsuperscript{b}, Babak Nadjar Araabi\textsuperscript{a} \\ \\
  \textsuperscript{a}School of Electrical and Computer Engineering,
  College of Engineering,
  University of Tehran,
  Tehran, Iran\\
  \textsuperscript{b}School of Medicine,
  Tehran University of Medical Sciences,
  Tehran, Iran \\ \\
  \textsuperscript{*}Corresponding Author \\
  Email address: b.khodabandehloo@ut.ac.ir
}

\begin{document}
\maketitle

\begin{abstract}
The brain's complex functionality emerges from network interactions that go beyond dyadic connections, with higher-order interactions significantly contributing to this complexity. One method of capturing higher-order interactions is through traversing the brain network using random walks.  The efficacy of these random walks depends on the defined mutual interactions between two brain entities. More precise capture of higher-order interactions enables a better reflection of the brain's intrinsic neurophysiological characteristics. One well-established neurophysiological concept is Homotopic Functional Connectivity (HoFC), which illustrates the synchronized spontaneous activity between corresponding regions in the brain's left and right hemispheres. We employ node2vec, a random walk node embedding approach, alongside resting-state fMRI from the Human Connectome Project (HCP) to obtain higher-order feature vectors. We assess the efficacy of different functional connectivity parameterizations using HoFC. The results indicates that the quality of capturing higher-order interactions largely depends on the statistical dependency measure between brain regions. Higher-order interactions defined by partial correlation, better reflects HoFC compare to other statistical associations. In this case of first-order interactions, tangent space embedding more effectively demonstrates HoFC. The findings validate HoFC and underscore the importance of functional connectivity construction method in capturing intrinsic characteristics of the human brain. 
\end{abstract}


\section{Introduction}
The brain's intricate functionality emerges from the collective interactions among its various components, and a deeper understanding of this interplay enhances our understanding of this complexity \cite{sporns2012discovering} , \cite{bassett2017network}, \cite{sporns2013structure}, \cite{sporns2016networks}. Consequently, many studies have adopted a network approach to further investigate the brain's underlying mechanisms \cite{van2010exploring}, \cite{bassett2011understanding}. A key element in analyzing the brain from a network perspective is the method of brain network construction. Typically, these networks are created and analyzed across three scales: micro, meso, and macro \cite{betzel2017multi}, \cite{sporns2016connectome}. At each scale, specific entities (nodes) and their interactions (edges) are defined based on the data modality, forming the respective brain network. It is critical that the definition of these nodes and edges accurately reflects the brain’s inherent neurophysiological characteristics \cite{bassett2017network}, \cite{dadi2019benchmarking} to ensure the validity of subsequent analyses conducted using statistical or machine learning techniques. 

Functional magnetic resonance imaging (fMRI) is a widely used modality to study the brain, and has revolutionized our understanding of the human brain function, enabling researchers to visualize and investigate brain activity with high precision \cite{ogawa1990brain}. Of particular interest in neuroimaging research is resting-state fMRI (rsfMRI), where spontaneous fluctuations in the brain’s activity are observed in the absence of any external stimuli \cite{buckner2013opportunities}, \cite{raichle2001default}, \cite{biswal1995functional}. Resting-state functional connectivity (FC) is derived from rsfMRI using statistical dependency measures between the activity pattern of different brain regions. The resulting FC provides a comprehensive depiction of how different brain areas interact at macro scale, leading to insights into the functional organization of the brain and its potential alterations in various neurological and psychiatric disorders \cite{van2010exploring}, \cite{fox2010clinical}, \cite{bassett2018understanding}, \cite{mulders2015resting}. 

A pivotal element of rsfMRI analysis involves assessing FC. This evaluation is critical, as FC matrices form the basis for further analysis with machine learning and statistical techniques, with the expectation that these matrices accurately mirror the brain's neurophysiological characteristics \cite{dadi2019benchmarking}, \cite{sanchez2021combining}. Among these characteristics, homotopic functional connectivity (HoFC) \cite{lowe1998functional}, \cite{salvador2005neurophysiological}, \cite{zuo2010growing} —the strong correlation between activities of corresponding brain regions across hemispheres— stands out as a fundamental and widespread characteristic, observed across humans and mammals \cite{salvador2005neurophysiological}, \cite{shen2015stable}, and notably affected in various brain diseases and disorders \cite{tang2016decreased}, \cite{hermesdorf2016major}. Despite its significance, the exploration of HoFC through machine learning methods remains limited, with most existing studies focusing on statistical analysis at the region of interest (ROI) or voxel level, highlighting a gap in the application of advanced analytical techniques to study this key neurophysiological characteristic \cite{jin2020functional}.

Machine learning techniques are widely employed in FC analyses \cite{vergun2013characterizing}, \cite{khosla2019machine}, \cite{du2018classification}. The key element of these techniques lies in extracting informative features from data. These features are then utilized in downstream tasks to draw conclusions. Feature vectors in FC can be extracted by treating each row as the feature vector for that specific region, capturing first-order interactions. However, the potential to derive even more informative feature vectors exists by exploring higher-order interactions through unsupervised representation learning techniques. The fundamental concept of these techniques is to extract more informative feature vectors from the data. Since FC is a graph, graph representation learning techniques seems to be appropriate choices to be applied to FC to obtain more informative feature vectors. One of the main techniques in graph representation learning is random walk (RW) node embedding, and node2vec \cite{grover2016node2vec} is one of the commonly used RW node embedding algorithms. Inspired by the word2vec model \cite{mikolov2013efficient}, this is an unsupervised representation learning algorithm on graphs \cite{goyal2018graph} and encode higher-order interactions by performing random walks on network connectivity. The advantage of using node2vec for representation learning on FC compared to other methods, such as Graph Neural Networks (GNNs), is that node2vec utilizes information encoded in the structure of the graphs, which is particularly relevant in the case of FC. In contrast, GNNs require node features that are not straightforward to define in brain networks \cite{wu2020comprehensive}. In practice, this algorithm generates an embedding vector for each node, which serves as the learned feature vector for that node. 

While dyadic interactions encompass a wealth of information, recent studies indicate that they might not fully capture interactions among brain regions \cite{rosenthal2018mapping}, \cite{levakov2021mapping}. Consequently, a fundamental question arises: Do higher-order interactions, captured through random walks on FC, contain more information than dyadic connections (first-order interactions) between two regions? If so, they should also better represent the intrinsic organizational characteristics of the human brain, such as HoFC, compared to first-order interactions. Some previous studies have shown the superiority of considering higher-order interactions in encoding brain structural connectivity \cite{levakov2021mapping}, \cite{glasser2016human}. Additionally, many studies have utilized node2vec for representation learning on FC \cite{chauhan2023classification}, \cite{rahimiasl2021random}, \cite{gan2023computer}, \cite{lama2022classification}. However, they have not investigated its superior performance compared to first-order relations.

In this study, we investigate the extent to which the FC (first-order interactions), obtained from various parameterizations, and its embedding matrix FCE (higher-order interactions), obtained using a RW node embedding approach, reflect HoFC. Our investigation is guided by four primary objectives: (1) To examine whether machine learning is capable of validating HoFC by using FC and FCE (2) To identify which FC parameterization better reflects HoFC and possesses greater neurobiological significance (3) To assess whether considering higher-order interactions offers advantages over first-order interactions in reflecting HoFC. (4) To evaluate which FC parameterization, when subjected to random walks, is more meaningful and generates more informative feature vectors. Essentially, to pinpoint the FC parameterization that, through the lens of statistical dependencies between two regions, allow for a more meaningful traversal of the graph via random walks.

\section{Method}
\label{sec:headings}

\begin{figure}[!ht]
    \centering
    \includegraphics[width=1.0\linewidth]{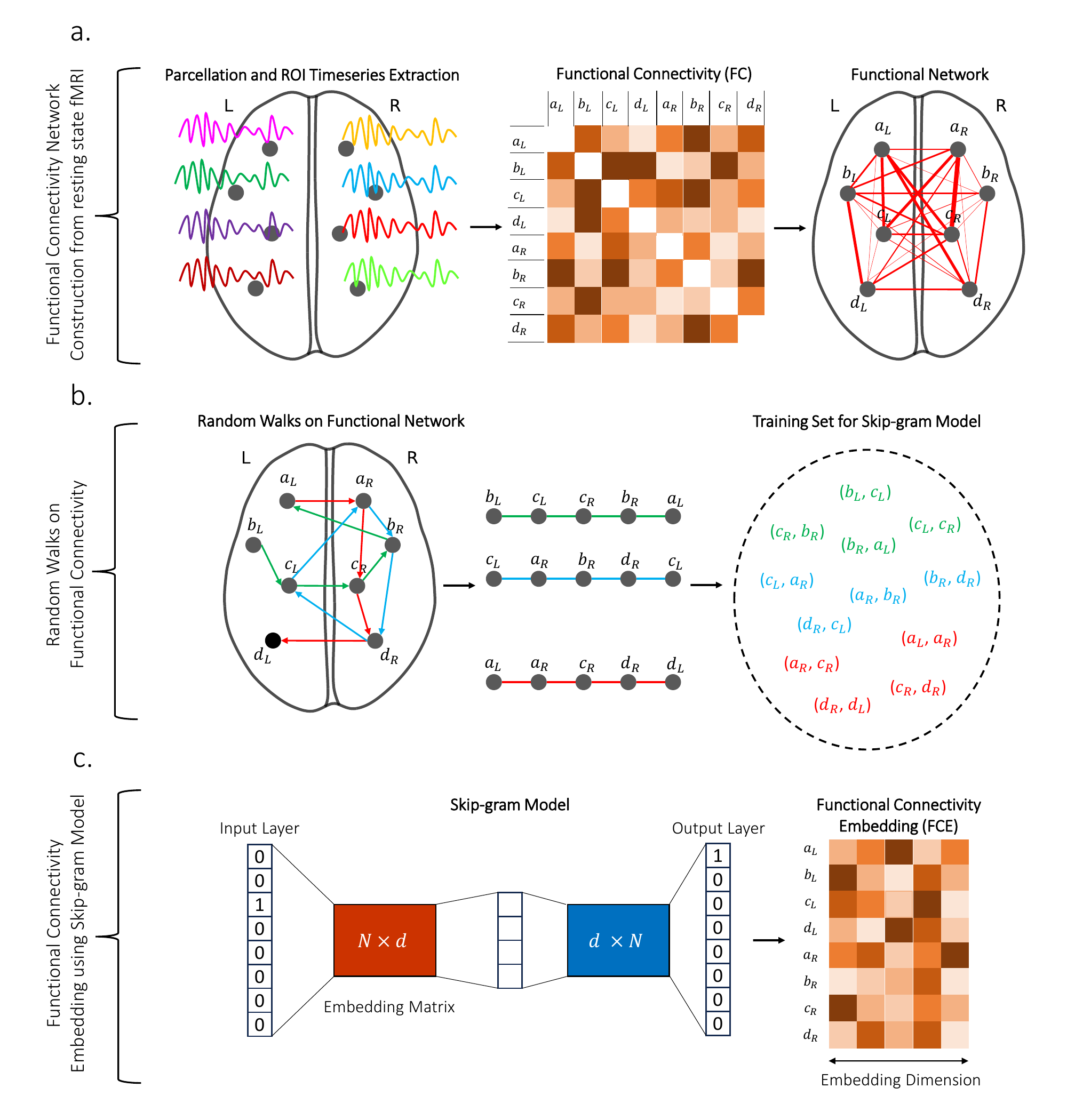}
    \caption{FC embedding workflow. a.  Functional connectivity (FC) is constructed from parcellated resting-state fMRI time series using different parameterizations such as bivariate correlation, partial correlation, or tangent space embedding  b. Node sequences are generated by performing random walks on the FC network. Subsequently, samples are created from these node sequences to form a training set c. The skip-gram model is trained using the generated training set to produce Region of Interest (ROI) embeddings (FCE). }
    \label{fig:fig1}
\end{figure}

\subsection{Dataset}
The analysis was conducted using resting-state fMRI time series of 100 independent subjects from human connectome project (HCP).  Comprehensive information regarding the HCP dataset is already available in prior publications \cite{van2012human}, \cite{glasser2013minimal}, \cite{smith2013resting}.The HCP functional pipeline was employed for data processing based on methods reported in previous publications \cite{glasser2013minimal}, \cite{smith2013resting}. The primary procedures included: initial spatial processing in both volumetric and gray ordinate formats, which involves representing brain locations as surface vertices \cite{glasser2013minimal}. This was followed by applying a weak high pass temporal filter with a full width at half maximum greater than 2000s to both data types for the elimination of slow drifts. For the volumetric data, MELODIC ICA \cite{routier2019new} was utilized, and artifacts were identified via FIX \cite{salimi2014automatic}. To eliminate artifacts and motion influences, regression was performed using several parameters: the six rigid-body parameter time-series, their temporal derivatives looking backward, and the squared values of all 12 regressors, applied to both volumetric and gray ordinate data \cite{glasser2013minimal}.

Time courses for each voxel underwent Z-scoring and subsequent averaging across brain regions. This was done for the 68 regions defined in the Desikan-Killiany atlas (excluding corpus callosum) \cite{desikan2006automated}, as well as the 200 regions outlined in the recent Yan 2023 \cite{yan2023homotopic} atlas. Any outlier time points that fell beyond three standard deviations from the mean were excluded. These two parcellation atlases were specifically chosen for hypothesis testing because they both had homotopic nature, having parallel parcels on each hemisphere. This processing was carried out using the Workbench software, specifically employing the 'workbench command-cifti-parcellate' \cite{marcus2011informatics}. Time-series for two sessions with two phase encoding directions (LR and RL) were merged for each subject using 'workbench command-cifti-merge'.

\subsection{Functional Connectivity Construction}
The Ledoit-Wolf estimator was chosen for covariance matrix estimation due to its superior ability to reduce noise and errors compared to the maximum likelihood estimator \cite{brier2015partial}, \cite{ledoit2004well}. To assess statistical associations between ROIs time series, three parameterizations were utilized: Pearson bivariate correlation, partial correlation, and tangent space embedding (TSE). Bivariate correlation, a widely used method for constructing FC, has limitations in brain network analyses due to potential confounding by other variables \cite{pearl2009causal}, \cite{reid2019advancing}. Partial correlation addresses this by isolating the relationship between two nodes, controlling for the influence of additional nodes, thus clarifying genuine correlations \cite{smith2013functional}, \cite{varoquaux2010brain}, \cite{smith2011network}. TSE, a manifold learning technique advantageous for FC analysis in brain imaging, transforms covariance matrices into a tangent space, simplifying the analysis of high-dimensional relationships and highlighting nuanced brain connectivity patterns \cite{dadi2019benchmarking}, \cite{brier2015partial}, \cite{islam2018multiband}, \cite{varoquaux2013learning}. For each parameterization, group-level analysis was conducted by generating a group average FC matrix from the FCs across subjects. All FC construction calculations were performed using the NiLearn Package \cite{abraham2014machine}.

\subsection{Network Embedding}

Node2vec \cite{grover2016node2vec}, an unsupervised learning algorithm for graph representation learning \cite{goyal2018graph}, extends the Word2Vec \cite{mikolov2013efficient} model's concept of word embeddings to graph nodes. By employing random walks to generate node sequences, node2vec captures both the local and global information within a graph, crucial for analyzing complex networks such as FC. The outcome of the node2vec algorithm is a feature vector for each node represented as a low-dimensional vector, reflecting its contextual relevance and neighborhood structure. For each parametrization, the node2vec is applied to the group-averaged FC to obtain node embeddings. Binarization of the networks is avoided to prevent information loss, and negative edges are ignored in random walks. Similar to the previous works \cite{rosenthal2018mapping}, \cite{levakov2021mapping}, we conducted 800 random walks from each node (rw = 800) with a walk length of 20 (l = 20). We set the embedding dimension to 30 (d = 30), the window size to 3 (ws = 3), and adjusted the neighborhood search parameters to p=0.1 and q=1.6. The process of applying node2vec to FC to obtain FCE is illustrated in Figure \ref{fig:fig1}. We utilized the PecanPy package implementation of node2vec \cite{liu2021pecanpy}, and all visualizations were created using the Matplotlib library \cite{hunter2007matplotlib}.

\subsection{Inter-hemispheric Analogy Test}

The Inter-hemispheric analogy test \cite{rosenthal2018mapping} was utilized  to evaluate the HoFC. In previous studies, this test has been employed to demonstrate the superiority of higher-order structural connectivity over first-order ones \cite{rosenthal2018mapping}, as well as to assess the quality of embedding alignment across individuals \cite{levakov2021mapping}. These analyses have been performed on structural connectivity created using Diffusion Tensor Imaging. The fundamental concept of this test posits that the relationship between each pair of regions in one hemisphere should mirror the pairwise relationship in the opposing hemisphere \cite{rosenthal2018mapping}.  The Examination of each pairwise relation is termed as an analogy. Therefore, the inter-hemispheric analogy test comprises $\binom{n}{2}$ analogies, where n represents the number of regions in one hemisphere. More precisely, each analogy computes similarity between a region and all other regions by considering one of its connections. For instance, consider that we want to assess HoFC for region named $A$. An analogy investigates the similarity between feature vector $\overrightarrow{A_{R}} - \overrightarrow{B_{R}} + \overrightarrow{B_{L}}$ ($B$ can be any other region) and all other region feature vectors.  Based on the HoFC concept, the similarity value for $\overrightarrow{A_{L}}$ (expected node) should be highest.  This test actually assesses the HoFC between $\overrightarrow{A_{R}}$ and $\overrightarrow{A_{L}}$. For each analogy, the cosine similarities of all other regions are arranged in the descending order. subsequently, the rank of the expected region (Here $\overrightarrow{A_{L}}$) is recorded. The lower the rank of this region, the better the HoFC is represented in an analogy. Similar to bin size chosen in \cite{rosenthal2018mapping}, for all subsequent statistical evaluations, A bin size of 5 for interhemispheric analogies were chosen in order to address the ambiguities of the HoFC spectrum, the noise in fMRI data, and the uncertainties of the node2vec algorithm.

Word2vec produces different outputs in different iterations \cite{smith2017offline}, \cite{dev2021closed}, \cite{wang2020towards}, and thus Node2vec also yields varying results in independent runs, attributable both to the uncertainty of random walks and the skip-gram model \cite{levakov2021mapping}. Consequently, we ran this algorithm 100 times on the group-averaged FC and conducted the inter-hemispheric analogy test in each run. The final desired rank for each analogy is determined as the median of these 100 ranks. A more detailed description of interhemispheric analogy test  and the procedure can be found in \cite{rosenthal2018mapping}. 

\section{Results}

\subsection{HoFC Validation using Machine Learning Techniques}

\begin{figure}[!ht]
  \includegraphics[width=\linewidth]{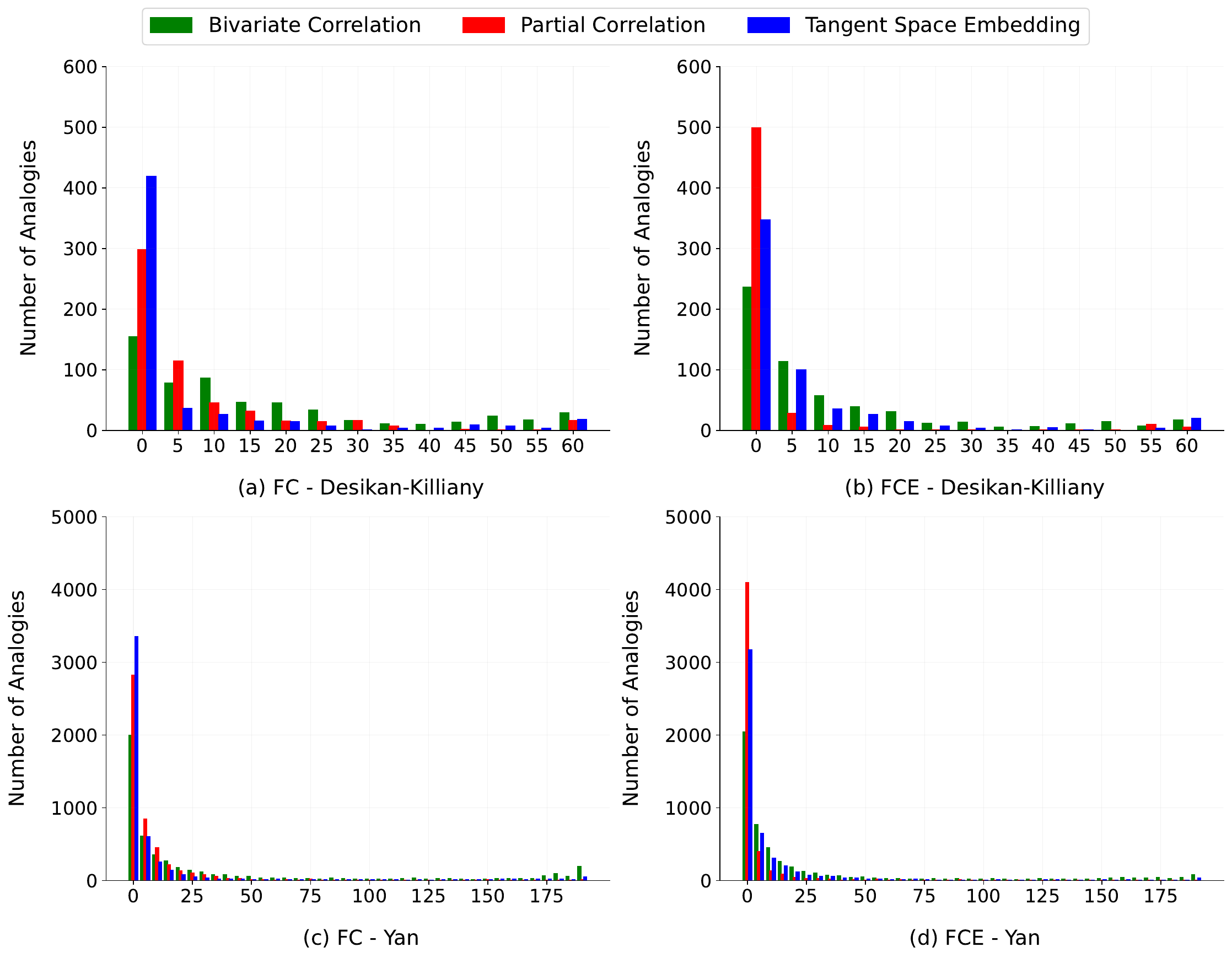}
  \caption{Performance of FC and FCE feature vectors in representing HoFC using interhemispheric analogies test. The ranking of the expected nodes was binned into bins of 5. For instance, the first column (consisting of three methods) shows the number of analogies in which the expected nodes were among the top 5 similar nodes. As it is observed in the figure the histograms have a right skewness validating HoFC visually. (a) FC - Desikan-Killiany. TSE is the most effective method in representing HoFC. (b) FCE - Desikan-Killiany. PC is the most effective method in representing HoFC. (c) FC - Yan. TSE is the most effective method in representing HoFC (d) FCE - Yan. PCs outperform others.}
  \label{fig:2}
\end{figure}

We first examined HoFC using machine learning feature vectors derived from either FC or FCE. Figure \ref{fig:2} displays histograms of inter-hemispheric analogies using both the Desikan-Killiany and Yan atlases. These histograms consistently show a rightward skewness, indicating that in interhemispheric analogies, the expected nodes were consistently ranked among the top five. We compared these results with those obtained from a series of randomized network ensembles, utilizing three distinct random network models \cite{vavsa2022null}. The initial model generated completely random adjacency matrices. The subsequent two null models were developed based on the original group-averaged functional connectivity. Specifically, the first null model involved reshuffling the weights within the original network, while the second null model was constructed to preserve the degree of each node, with weights for each node being generated under this constraint, although complete preservation was not always achievable. 

In the case of FC, for a specific type of random network model, we generated 100 random networks from group-averaged functional connectivity. We compared the distributions of analogies for each network generated random network against the distribution of the original group's average functional connectivity, employing the Kolmogorov-Smirnov statistical test \cite{massey1951kolmogorov}. This analysis was conducted using both the Desikan-Killiany and Yan atlases. For each atlas and across the three null network models, more than 90 out of 100 comparisons demonstrated statistically significant differences, with several instances where all 100 out of 100 comparisons showed statistically significant differences. This validates HoFC using first-order interactions.  

For FCE, we applied node2vec ten times to each of the 100 random networks generated previously and averaged the results to address the inherent uncertainty of the node2vec algorithm, which can yield different outcomes across iterations. As a result, we have 100 averaged FCEs for each random network model. Analogy tests were conducted for each, producing 100 histograms. We also performed node2vec ten times on the original group-averaged functional connectivity, averaged these results, and then conducted inter-hemispheric analogy tests to generate a histogram for the original data. These histograms were then compared to those from the random network models using the Kolmogorov-Smirnov test, revealing that in all cases involving random network models, more than 90 out of 100 comparisons yielded statistically significant differences.

\subsection{Assessment of FC and FCE using HoFC}

\begin{figure}[!ht]
  \includegraphics[width=\linewidth]{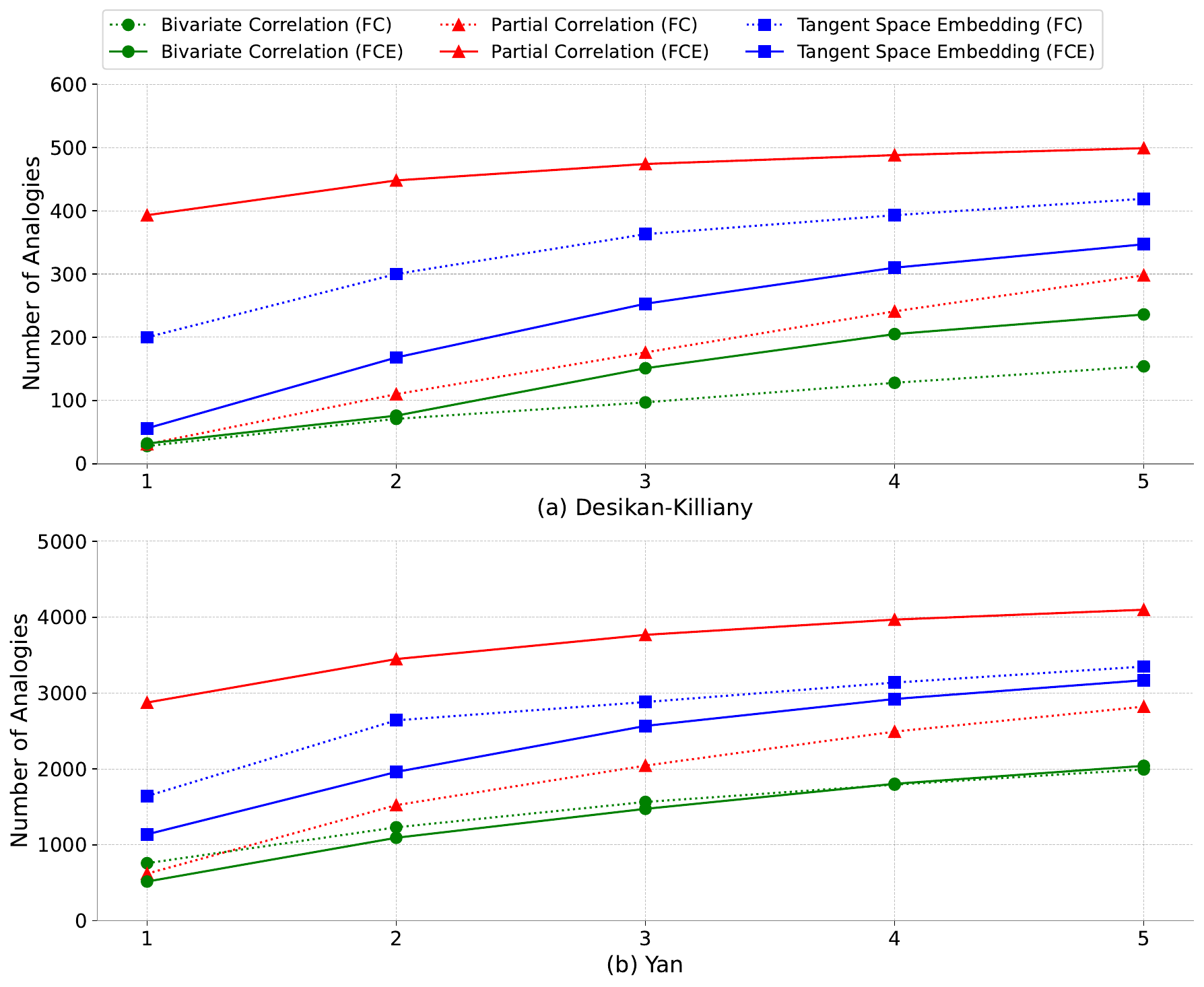}
  \caption{Interhemispheric analogies test. (A) Yan. In FC, TSE outperforms other methods in reflecting HoFC, and
no significant or consistent difference is observed between bivariate correlation and partial correlation. In the case of FCE, there is a consistent and statistically significant difference among all three methods, with partial correlation outperforming the others. Higher-order relations in TSE outperform first-order relations in reflecting HoFC. No consistent superiority is observed in bivariate correlation in the case of FC and FCE. Overall, two significant patterns are observed: 1. Partial correlation with intervening variable control best reflects HoFC; 2. TSE in the case of first-order relations better reflects HoFC. (B) Desikan-Killiany. In FC, TSE outperforms other methods, and no significant or consistent result is observed between bivariate correlation and partial correlation. In FCE, a significant
and consistent difference is observed among the three methods, with partial correlation outperforming the others.
When comparing FC and FCE, partial correlation is more advantageous in FCE. However, TSE performs better in
FC compared to FCE. No significant difference is observed in bivariate correlation between FC and FCE.}
  \label{fig:3}
\end{figure}

After validating HoFC using FC and FCE, our analysis focused on comparing different parameterizations of FC using HoFC. We considered analogy values within the range of [0, 5), as demonstrated in Figure \ref{fig:3}. In this representation, a bin size of 1 denoted an expected rank within the range [0, 1), while a bin size of 2 corresponded to expected ranks in the range [1, 2), and so forth. The analysis of expected node ranks within [0, 5) is critical since the precise similarity range for HoFC is unknown \cite{rosenthal2018mapping}, \cite{jin2020functional}. Hence, in order to draw a conclusive inference, it is imperative to ascertain the consistency of analogy test results across these diverse bin sizes. Additionally, placing exclusive emphasis on the top-ranking node might prove excessively restrictive. Our initial step was to identify consistent results across different bin sizes and atlases. Upon finding consistent results, we proceeded with further statistical analysis. Our analysis involved several comparative evaluations: initially, we investigated analogies derived from three statistical association methods and considering first order relations (FC comparison). We then examined analogies resulted from considering higher order relations (FCE comparison). Finally, we conducted a comparison between FC and FCE. Final results are summarized in Table \ref{demo-table}. 

\subsubsection{Inter-hemispheric Analogy Test for FC}

In the case of first-order feature vectors (FC), TSE outperformed both bivariate and partial correlation methods in reflecting the HoFC across various bin sizes, as shown in Figure \ref{fig:3}, within both the Yan and Desikan-Killiany atlases. Despite the inconsistent performance between bivariate and partial correlations, TSE consistently demonstrated superior efficacy. In the Yan Atlas, out of 4,950 analogies, TSE ranked the expected node within the top five positions 3,349 times, significantly more than bivariate (1,565 times) and partial correlation (2,821 times). This was statistically supported by chi-square tests, revealing significant differences in performance between TSE and bivariate correlation $({\chi}^2(1, N=9900) = 1285.98, p<0.001)$, and TSE and partial correlation $({\chi}^2(1, N=9900) = 119.92, p<0.001)$. Similarly, in the Desikan-Killiany, out of 561 analogies, TSE's performance was superior, ranking the expected node in the top five 419 times, compared to 154 for bivariate and 298 for partial correlation, with chi-square tests confirming significant differences (bivariate vs. TSE: $({\chi}^2(1, N=1122) = 248.58, p<0.001)$; partial correlation vs. TSE: $({\chi}^2(1, N=1122) = 55.64, p<0.001)$. 

\subsubsection{Inter-hemispheric Analogy Test for FCE.}

In the case of higher-order feature vectors (FCE), partial correlation and TSE demonstrated superior capabilities in reflecting the HoFC across bin sizes, outshining bivariate correlation in both the Yan and Desikan-Killiany atlases. Specifically, in the Yan atlas analysis involving 4,950 analogies, partial correlation led by ranking the expected node within the top five in 4,098 cases, followed by TSE with 3,169 instances, and bivariate correlation at 2,041. Chi-square tests underscored these differences, showing significant statistical disparities: bivariate versus partial correlation $({\chi}^2(1, N=9900) = 1812.51, p<0.001)$, bivariate versus TSE $({\chi}^2(1, N=9900) = 514.60, p<0.001)$, and partial correlation versus TSE $({\chi}^2(1, N=9900) = 445.58, p<0.001)$, with a comprehensive analysis further highlighting the variance in method performance $({\chi}^2(2, N=14850) = 1832.80, p<0.001)$. In the Desikan-Killiany atlas, from 561 analogies, partial correlation again topped with 499 top-five rankings, TSE followed with 347, and bivariate correlation was at 236, with significant chi-square results illustrating differences among the methods (bivariate vs. partial: $({\chi}^2(1, N=1122) = 270.77, p<0.001)$; bivariate vs. TSE: $({\chi}^2(1, N=1122) = 43.20, p<0.001)$; partial vs. TSE: $({\chi}^2(1, N=1122) = 109.56, p<0.001)$, and a collective test confirming significant performance variations $({\chi}^2(2, N=1683) = 270.70, p<0.001)$. 

\subsubsection{Functional Connectivity Embedding (FCE) vs Functional connectivity (FC)}

In assessing the effectiveness of FCE versus FC in capturing HoFC, it was observed that in both atlases, partial correlation in FCE significantly outperformed its counterpart in FC. Conversely, for TSE, the FC method demonstrated superior representation of HoFC compared to FCE in both atlases. However, no significant or consistent trend was noted in the performance of bivariate correlation between FC and FCE. Specifically, in the Yan atlas, bivariate correlation in FC outperformed that in FCE, while in the Desikan-Killiany atlas, the reverse was true with bivariate correlation in FCE outperforming that in FC. In Yan atlas, a significant difference was found in partial correlation between FC (2821) and FCE (4098), yielding $({\chi}^2(1, N=9900) = 781.50, p < 0.001)$. For TSE FC (3349) compared to FCE (3169) showed a significant disparity $({\chi}^2(1, N=9900) = 14.39, p < 0.001)$. In Desikan-Killiany atlas, a significant difference was found in partial correlation between FC (298) and FCE (499), with $({\chi}^2(1, N=9900) = 173.27, p < 0.001)$. TSE also showed significant differences between FC (419) and FCE (347) $({\chi}^2(1, N=9900) = 20.74, p < 0.001)$.

\renewcommand{\arraystretch}{3.5}
\begin{table}[!h]
\begin{center}
\caption{Based on all analyses, this table presents and compares the results of FC and FCE. For FC, TSE was the
most effective method in reflecting HoFC, while bivariate correlation yielded inconsistent results, and partial correlation was effective but less so than TSE. In FCE, partial correlation emerged as the most effective method, surpassing the other two methods. In the comparison of FC versus FCE, partial correlation was more effective in FCE, while TSE was more effective in FC. No significant pattern was observed in bivariate correlation between FC and FCE. Overall, partial correlation in FCE and TSE in FC were found to be the most effective methods in capturing
HoFC, respectively. \\}
\label{demo-table}
\begin{tabular}{| c | c | c | c |} 
 \hline
 & \cellcolor{gray!25}Bivariate Correlation & \cellcolor{gray!25} Partial Correlation & \cellcolor{gray!25} \makecell{Tangent Space Embedding} \\
 \hline
  \cellcolor{gray!25} FC & Inconsistent results  & Less Effective than TSE  & Most Effective \\ 
 \hline
 \cellcolor{gray!25} FCE & Least Effective  & Most Effective  & \makecell{More Effective \\ than Bivariate}  \\
 \hline
 \cellcolor{gray!25} FC vs FCE & \makecell{No significant difference \\ between FC and FCE}  & \makecell{More effective in FCE than \\in FC, and most effective \\ method in both FC and FCE}   & \makecell{More effective in FC \\ than in FCE}  \\
 \hline
\end{tabular}
\end{center}
\end{table}

\subsection{ROIs Representing HoFC}

\begin{figure}[!ht]
    \centering
    \begin{subfigure}[b]{\textwidth}
        \centering
        \includegraphics[width=0.32\linewidth]{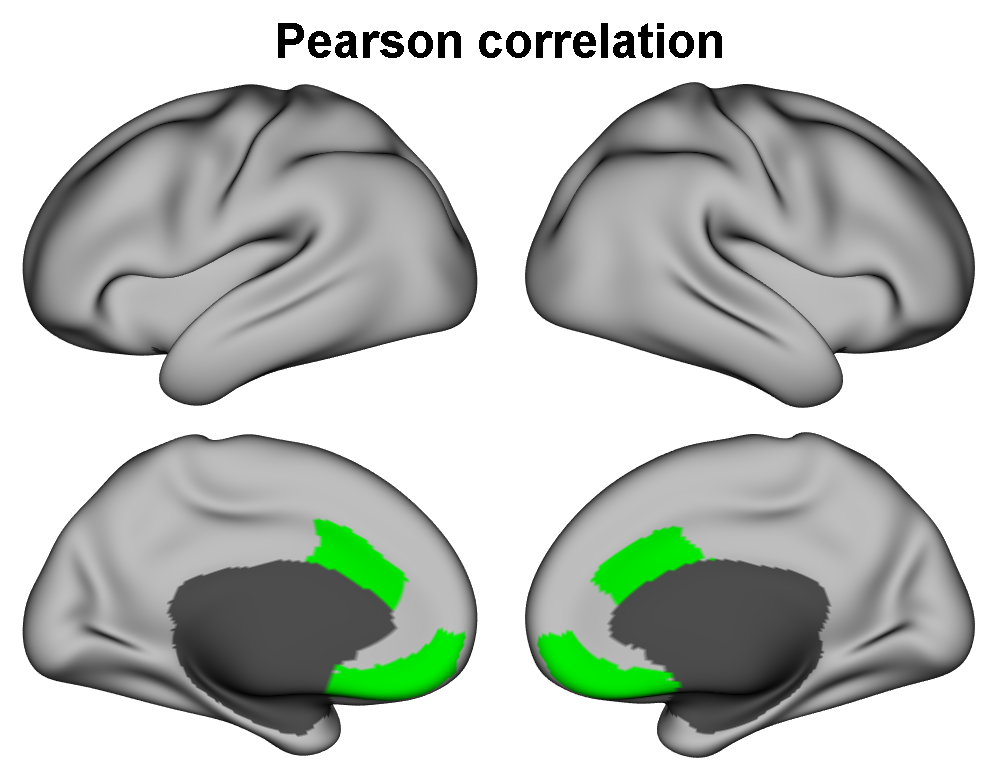}%
        \hfill
        \includegraphics[width=0.32\linewidth]{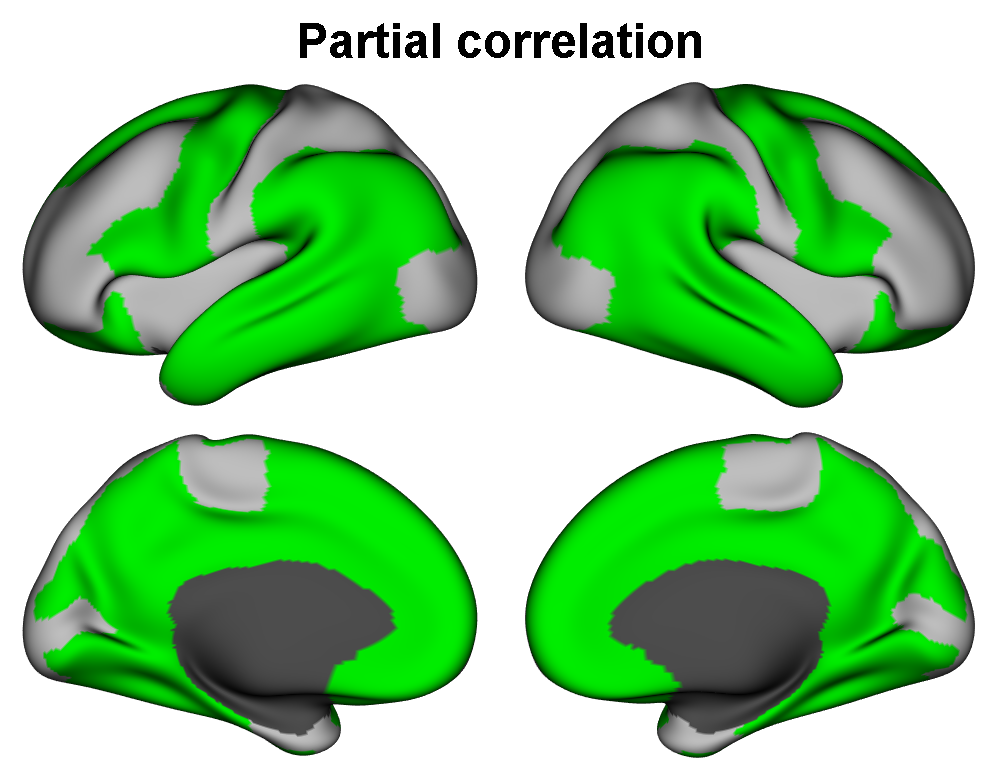}
        \hfill
        \includegraphics[width=0.32\linewidth]{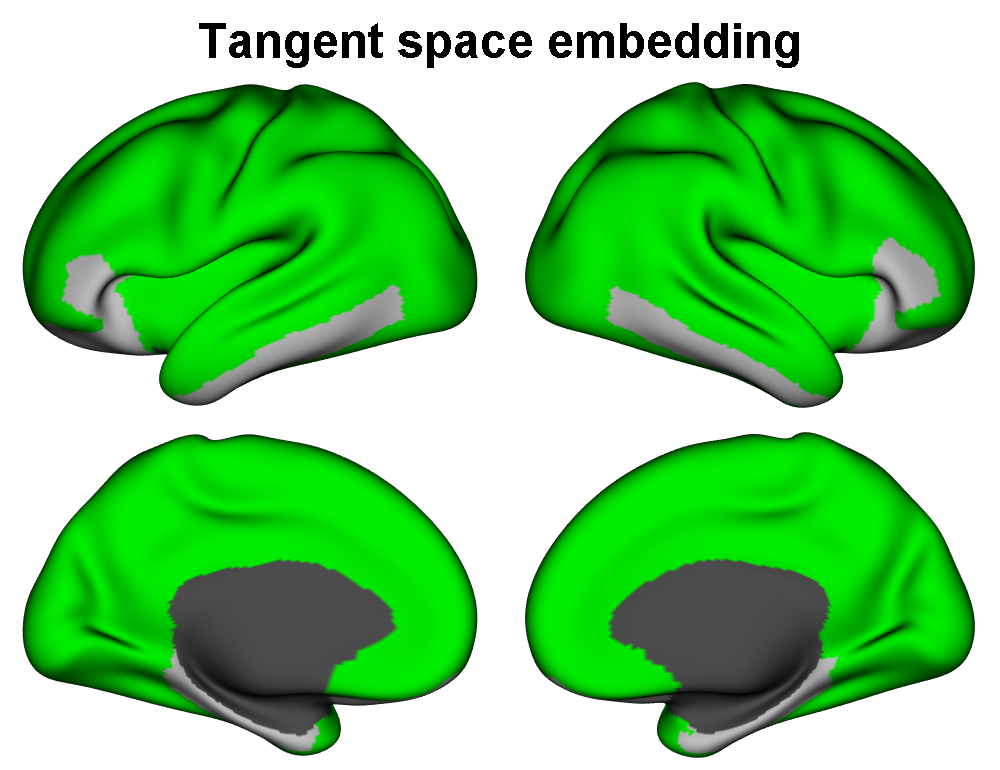}
        \caption{FC - Desikan-Killiany}
    \end{subfigure}
    \begin{subfigure}[b]{\textwidth}
        \centering
        \includegraphics[width=0.32\linewidth]{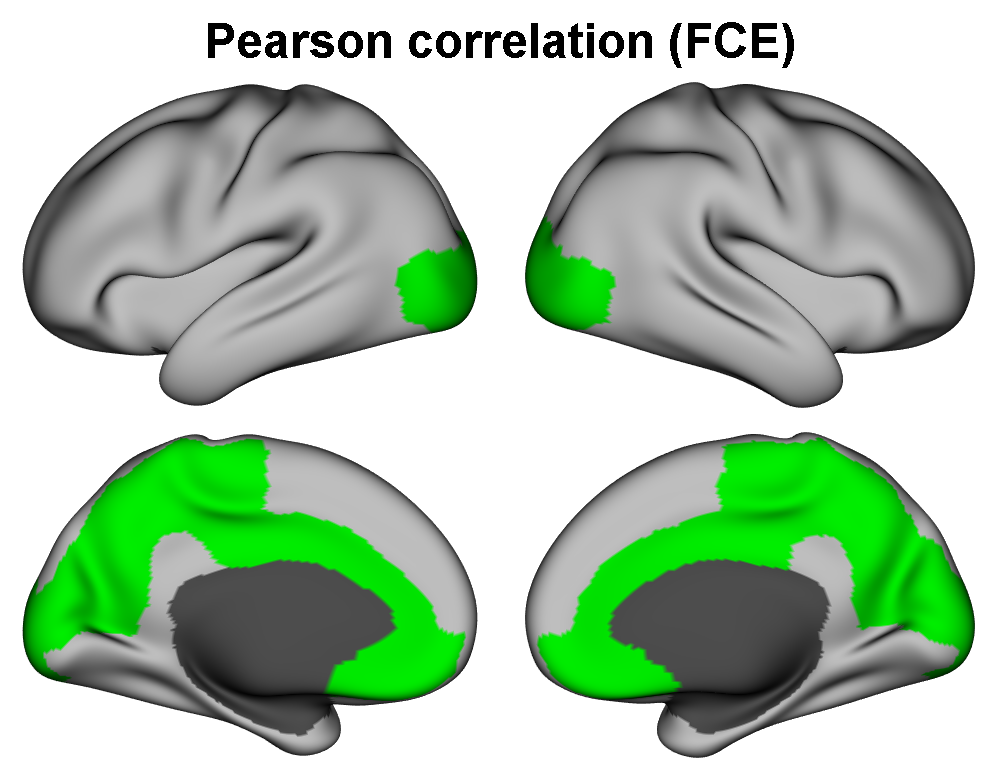}%
        \hfill
        \includegraphics[width=0.32\linewidth]{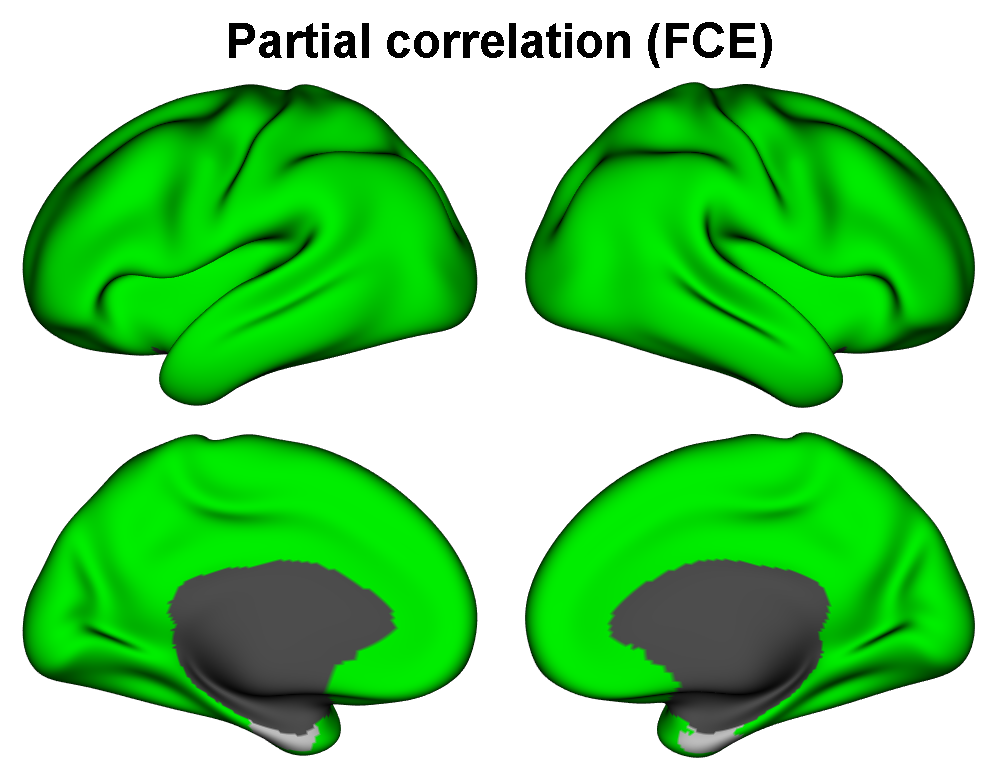}
        \hfill
        \includegraphics[width=0.32\linewidth]{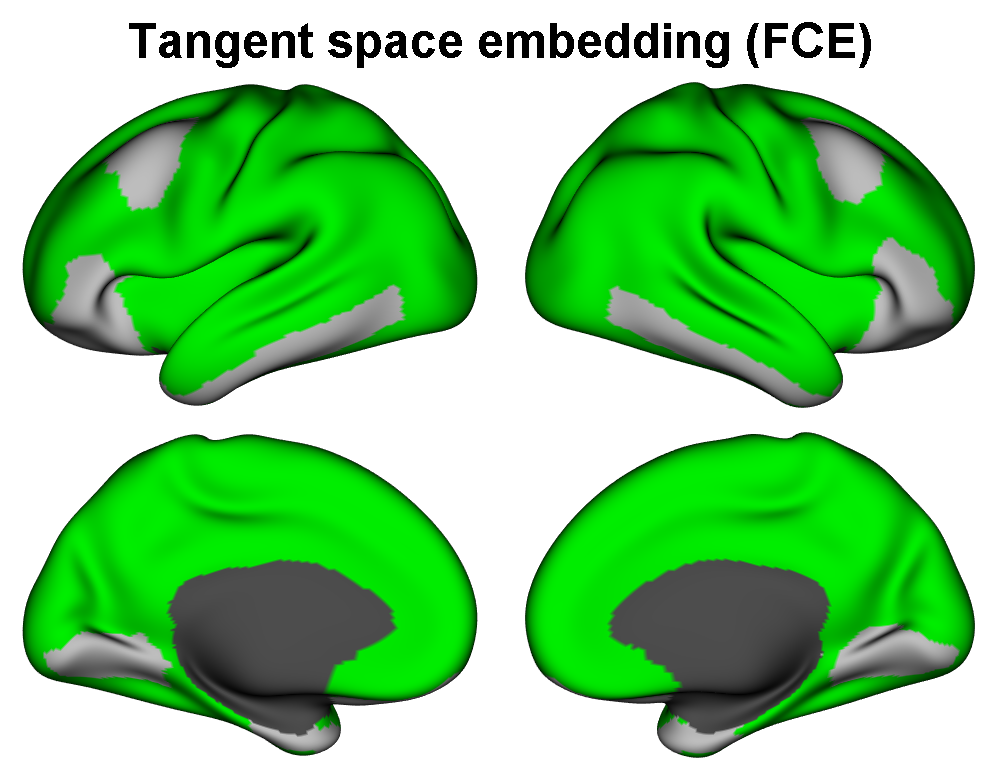}
        \caption{FCE - Desikan Killiany}
    \end{subfigure}
    \begin{subfigure}[b]{\textwidth}
        \centering
        \includegraphics[width=0.32\linewidth]{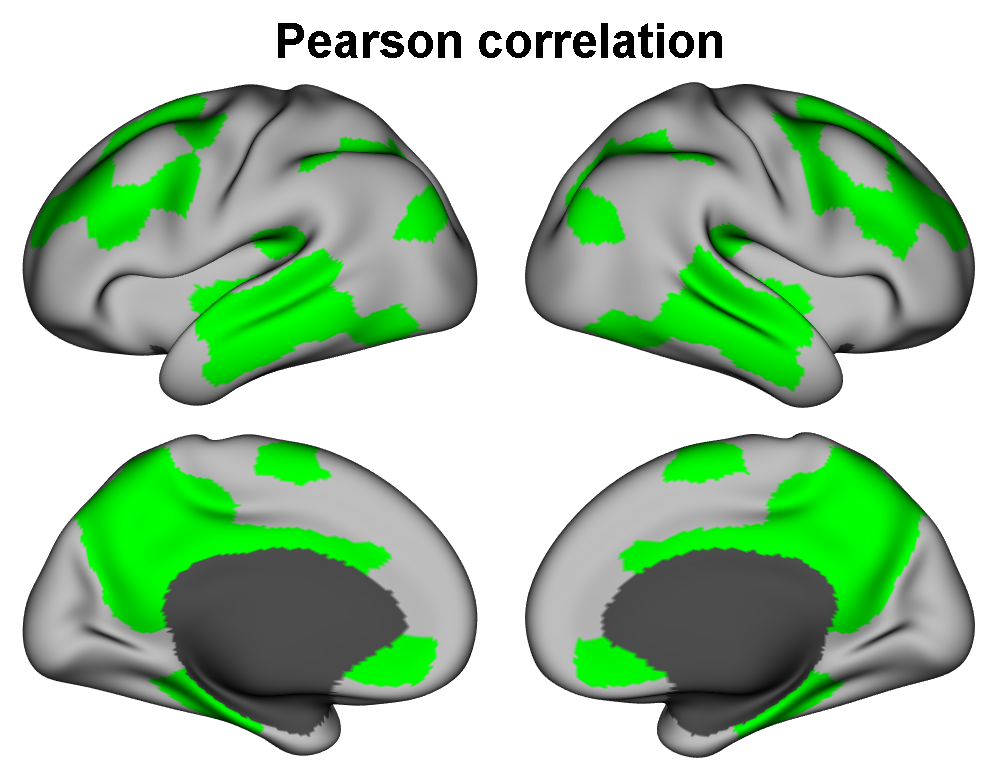}%
        \hfill
        \includegraphics[width=0.32\linewidth]{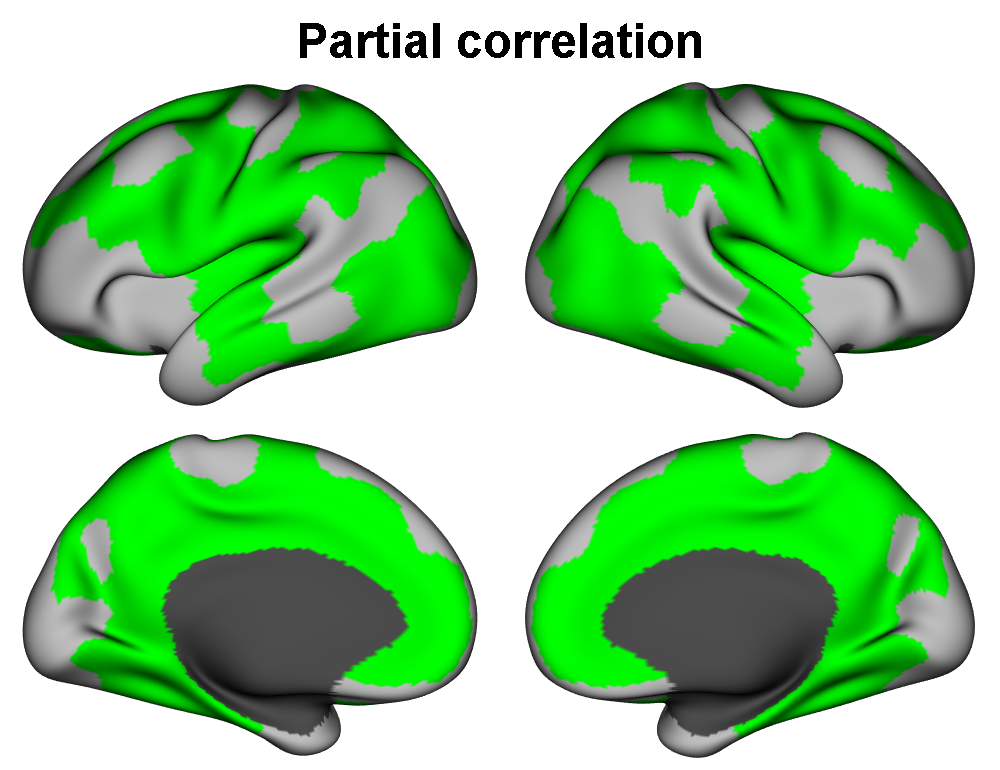}
        \hfill
        \includegraphics[width=0.32\linewidth]{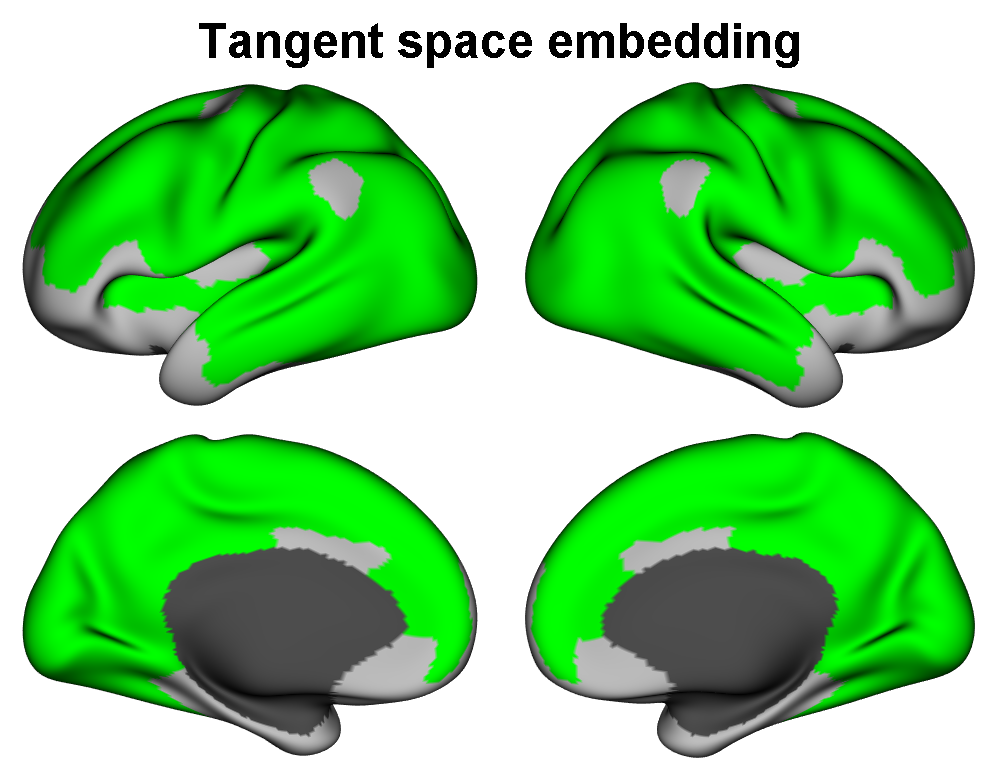}
        \caption{FC - Yan}
    \end{subfigure}
    \begin{subfigure}[b]{\textwidth}
        \centering
        \includegraphics[width=0.32\linewidth]{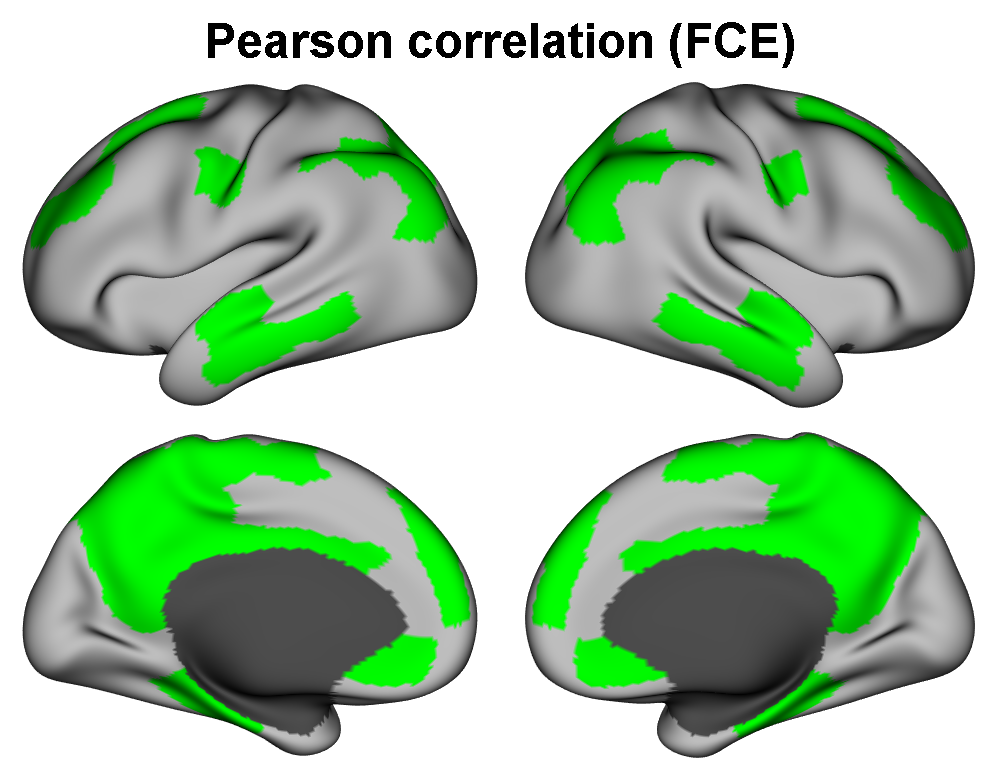}%
        \hfill
        \includegraphics[width=0.32\linewidth]{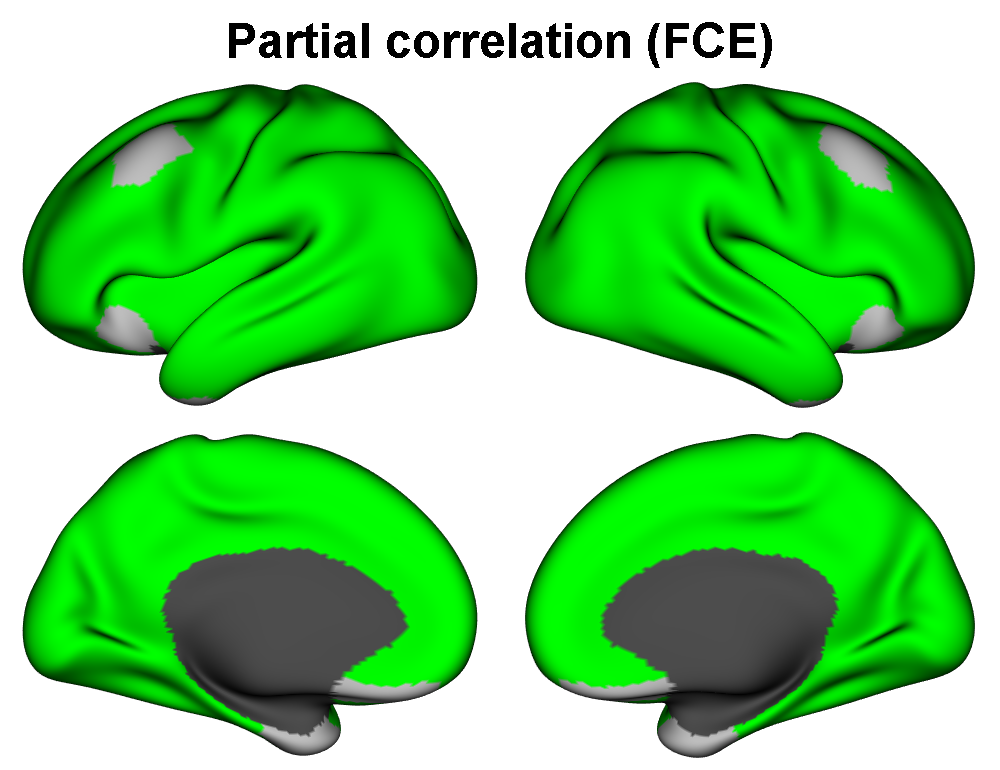}
        \hfill
        \includegraphics[width=0.32\linewidth]{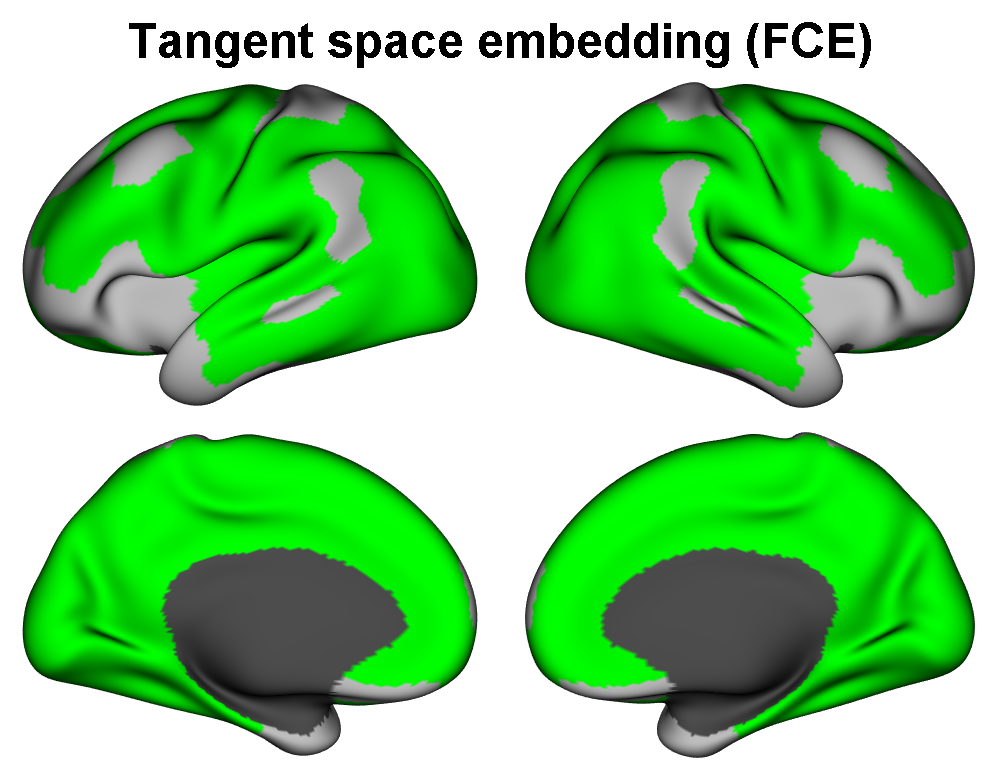}
        \caption{FCE - Yan}
    \end{subfigure}
   \caption{Comparison of regions between the Desikan-Killiany and Yan atlases. Regions where expected nodes ranked among the top five are highlighted in green. The first and second rows depict Functional Connectivity (FC) and Functional Connectivity Embedding (FCE) in the Desikan-Killiany atlas, respectively, while the third and fourth rows depict FC and FCE in the Yan atlas.}
    \label{fig:4}
\end{figure}

In addition to statistically examining HoFC through inter-hemispheric analogies, Figure \ref{fig:4} presents visualizations of the expected regions consistently ranked among the top 5. For each analogy, the ranks of both nodes are recorded. Subsequently, the median of these ranks for each region is computed, serving as the region rank. Regions with a rank within the range [0, 5) are then visualized in green. 

\section{Discussion}
Our study explores homotopic functional connectivity (HoFC), a significant characteristic of mammalian brain networks, by using first-order and higher-order feature vectors derived from resting-state functional interactions through machine learning techniques. Our key finding is that random walks on functional connectivity, created using partial correlation statistical dependency, more accurately capture the intrinsic neurophysiological characteristics of the human brain. This highlights the significance of both higher-order interactions and the choice of statistical dependency measure in constructing FC. It suggests that for a more effective encoding of FC using the random walk node embedding method, the statistical dependency measure that defines connections between two regions should focus exclusively on the dependency of their activations while excluding the influence of other nodes. Partial correlation accomplishes this by calculating a coefficient that more precisely represents the direct and genuine interaction between two nodes, mitigating the impact of other intervening variables. Consequently, our results indicate that the best outcomes are achieved using random walks on FC created using partial correlation as the weight between two nodes. 

We first validated HoFC using machine learning feature vectors by comparing to various network null models. The findings imply that the corresponding regions in opposite hemispheres possess the most similar feature vectors, as computed by cosine similarity. These feature vectors, which encompass both first-order and higher-order interactions, effectively validate HoFC. To the best of our knowledge, little work has been done on investigating HoFC using machine learning techniques.

In the case of first-order interactions (FC), tangent space embedding (TSE) better represents HoFC compared to other parametrizations. Previous results have shown that TSE results in higher classification accuracy relative to bivariate correlation and partial correlation \cite{dadi2019benchmarking}. These two findings are in accord with each other, as a method that better captures the intrinsic characteristics of the human brain can potentially perform better in machine learning tasks such as classification and regression. Indeed, TSE captures the correlation of neuronal activity across different brain regions by transforming them into a tangent space, thereby simplifying their complex, high-dimensional relationships. The difference between the effectiveness of TSE and other methods is more significant compared to the difference between bivariate correlation and partial correlation, generally indicating that first-order feature vectors derived using TSE better captures the neurophysiological characteristics of the human brain. Additionally, there does not seem to be a consistent, statistically meaningful difference between bivariate correlation and partial correlation.

In the case of higher-order interactions (FCE), partial correlation significantly and consistently outperforms other parameterizations in representing HoFC. One key element of RW node embedding approaches is the generation of node sequences, which serve as training samples. Thus, it is crucial that these node sequences are meaningfully generated, requiring that edge weights accurately reflect the genuine connections between nodes. To enhance the relevance of graph traversal, these values should be computed in a way that removes the effects of other nodes. Therefore, considering partial correlation as the statistical dependency measure might be more advantageous for generating random walks on FC. Conversely, it has been shown that bivariate correlation tends to produce substantial spurious or false positive edges \cite{sanchez2021combining}, which can mislead the random walks on graph connectivity. Moreover, if an edge truly exists, its value could be influenced by other nodes, making it less informative for reflecting direct and genuine interactions, a necessity for generating meaningful random walks. These factors explain why bivariate correlation performs poorly in FC embedding when demonstrating neurophysiological characteristics of the human brain, with TSE performing better than bivariate correlation but worse than partial correlation in the case of FCE.

When comparing first-order and higher-order interactions (FC vs. FCE) using bivariate correlation, no consistent or significant differences are observed in how well they reflect HoFC. More specifically, performance sometimes deteriorates with higher-order interactions compared to first-order interactions. Additionally, performance is sensitive to bin size. Overall, these findings do not support the superiority of higher-order interactions over first-order interactions in representing HoFC when using bivariate correlation. This may indicate that random walks on FC, created using bivariate correlation, fail to capture higher-order interactions, possibly due to the intrinsic limitations of the method \cite{reid2019advancing}, \cite{spirtes2001causation}. In the case of TSE, FC better represents HoFC compared to FCE. This suggests that random walks on the connectivity matrix constructed by TSE may not be as informative and fail to adequately capture HoFC. Indeed, TSE primarily involves feature learning on the covariance matrix and, in terms of correlation-based statistical dependencies, it less accurately represents true dyadic interactions between two nodes in a network. Consequently, performing random walks on TSE might be less meaningful. On the other hand, partial correlation more effectively represents HoFC and significantly outperforms other methods in capturing pure dyadic interactions between two nodes. This underscores that merely using a more advanced method, such as node2vec, may not adequately capture the intrinsic characteristics of functional connectivity. Indeed, the method of constructing FC is crucial because RW embedding approaches traverse the graph based on edge weights, which influences their path through the graph. Therefore, the manner in which these graphs are constructed is of paramount importance for subsequent analysis using machine learning methods.

\section{Conclusion}
In this study, we explore Homotopic Functional Connectivity (HoFC) by leveraging both first-order and higher-order feature vectors derived from three different parameterizations of the resting-state functional connectivity. Our validation of HoFC using machine learning techniques demonstrates that partial correlation is the most suitable framework for random walk node embeddings, and the corrponding encodings better capture intrinsic neurophysicalogical charactersitics of the human brain. Here,  We do not investigate other graph exploration parameters (p, q), which are crucial for encoding the graph based on homophily or structural roles. Understanding these parameters remains vital for improving FC encoding. Additionally, we do not examine the spectrum of HoFC within regions of interest (ROIs), identifying regions exhibiting HoFC using only a strict threshold. Looking forward, we aim to focus on how well partial correlation can predict individual differences in both healthy subjects and patients. This approach will be relevant for classification or regression problems. We also plan to adopt more sophisticated methods for establishing statistical associations in neural time series, which will enable deeper insights into HoFC based on other parameterizations.

\section*{Acknowledgments}
Data were provided [in part] by the Human Connectome Project, WU-Minn
Consortium (Principal Investigators: David Van Essen and Kamil Ugurbil; 1U54MH091657)
funded by the 16 NIH institutes and Centers that support the NIH blueprint for Neuroscience
Research; and by the McDonnell Center for Systems Neuroscience at Washington University.

\bibliographystyle{unsrt}  
\bibliography{references}

\end{document}